%
%
\font\medss=cmss10 scaled\magstephalf

\magnification \magstep1
\hsize=17.0 true cm
\vsize=25.0 true cm
\hoffset=-0.4 true cm
\voffset=-1.2 true cm
\baselineskip=12pt
\parskip=0.5\baselineskip
\parindent=2.5em
\input rotate
\input epsf \epsfverbosetrue
%
%
\font\bigbold=cmbx12 scaled\magstep1
\font\bigbold=cmbx12 
\font\medcaps=cmcsc10 scaled\magstep1
\font\medslant=cmsl9 scaled\magstep1

\font\smallbold=cmbx9
\font\smallital=cmti9
\font\smallroman=cmr9

%
%
\nopagenumbers
\def\moriondtitle#1{
	\vglue 4cm
	\centerline{
		\bigbold
		#1
	}
}
\def\moriondtitlewheader#1{
\noindent{\smallital XIVth Moriond Meeting, ``Clusters of Galaxies'',
ed. F. Durret, A. Mazure \& J. Tran Thanh Van (Gif-sur-Yvette: Editions
Fronti\`eres)} 
	\vglue 3.0 true cm plus 0pt minus 0pt
	\centerline{
		\bigbold
		#1
	}
\headline={
	\ifodd
		\pageno\rightheadline
	\else
		\leftheadline
	\fi
}
\def\rightheadline{
	\medss\hfil\folio
}
\def\leftheadline{
	\medss\noindent$\!\!\!\!$\folio\hfil
}
}
\def\moriondtitletwo#1{
	\vskip0.5\baselineskip
	\centerline{
		\bigbold
		#1
	}
}
\def\moriondauthors#1{
	\vskip 2\baselineskip
	\centerline{
		\medcaps
		#1
	}
}

\def\moriondinst#1{
	\centerline{
		\medslant
		#1
	}
}

\outer\def\moriondabstract\par{
	\baselineskip=12pt
	\vskip 5.5 true cm	
	\vskip0pt plus 1.0\baselineskip\penalty-250
	\vskip0pt plus -1.0\baselineskip
	\vskip1.0\baselineskip
	\leftline{\bf Abstract}\nobreak
	\vskip -0.2\baselineskip\noindent
}
\outer\def\moriondsec#1\par{
	\baselineskip=15pt
	\vskip0pt plus 1.0\baselineskip\penalty-250
	\vskip0pt plus -1.0\baselineskip
	\vskip1.0\baselineskip
	\leftline{\bf #1}\nobreak
	\vskip -0.2\baselineskip\noindent
}
\outer\def\moriondsubsec#1\par{
	\vskip0pt plus 1.0\baselineskip\penalty-250
	\vskip0pt plus -1.0\baselineskip
	\vskip0.6\baselineskip
	\leftline{\sl #1}\nobreak
	\vskip -0.3\baselineskip\noindent
}
\outer\def\moriondsubsecone#1\par{
	\vskip0pt plus 1.0\baselineskip\penalty-250
	\vskip0pt plus -1.0\baselineskip
	\vskip-1.0\baselineskip
	\leftline{\sl #1}\nobreak
	\vskip -0.3\baselineskip\noindent
}

%

%
%

%
%

%
%

%
%
\newcount\figureno
\def\moriondfiguretwo#1#2#3#4{
	\goodbreak
	\midinsert
		\vskip#2 true cm
		\noindent{\bf Figure #1.} #3
		\par
		\figureno=#1
		\advance\figureno by 1
		\noindent{\bf Figure \number\figureno.} #4
		\vskip-0.5\baselineskip
	\endinsert
}
%
%

%
%
\def\moriondpsfigurenr#1#2#3#4{
	\goodbreak
	\midinsert
		\leftskip=1 true cm
		\rightskip=1 true cm
		\setbox100=\hbox{\epsfysize=#2 true cm\epsffile{#3}}
		\vskip -0.5\baselineskip
		\centerline{\hbox{\box100}}
		\vskip -0.5\baselineskip
		{\baselineskip=12 pt
			\noindent{\bf Figure #1.} #4
		}
		\vskip-0.5\baselineskip
		\leftskip=0 true cm
		\rightskip=0 true cm
	\endinsert
}
%
%

%
%

%
%

%
\def\moriondrefhead{
	\vskip0pt plus 1.0\baselineskip\penalty-250
	\vskip0pt plus -1.0\baselineskip
	\vskip\baselineskip
	\parindent=0pt
	\parskip=0pt
	\vbox{
		\centerline{\bf References}
		\vskip0.5\baselineskip
	}
	\baselineskip=12pt
}
%
%
\def\moriondpaper#1#2#3#4#5#6{
	\hangindent=3em
	\hangafter=1
	\smallroman #6] #1, #2,
	{\smallital #3\/},
	{\smallbold #4}, #5.
	\par
}
%
%

%
%

%
%
\def\moriondinpress#1#2#3#4{
	\hangindent=3em
	\hangafter=1
	\smallroman #4] #1, #2,
	{\smallital #3\/},
	in press.
	\par
}
%
%
\def\moriondsubmit#1#2#3#4{
	\hangindent=3em
	\hangafter=1
	\smallroman #4] #1, #2,
	submitted to {\smallital #3\/}.
	\par
}
%
%

%
%

%
%

%
%

%
%

%
%

%
%

%
%
\def\moriondproceed#1#2#3#4#5#6#7#8{
	\hangindent=3em
	\hangafter=1
	\smallroman #8] #1, #2,
	{\smallital in #3\/},
	ed. #4 (#5: #6), p. #7.
	\par
}
%
%
\def\moriondproceedinpress#1#2#3#4#5#6#7{
	\hangindent=3em
	\hangafter=1
	\smallroman #7] #1, #2,
	{\smallital in #3\/},
	ed. #4 (#5: #6), in press.
	\par
}
%
%

%
%

%
%

%
%

%
%


%
\def \srule {\vskip 0.4\baselineskip \hrule height.7pt
 \vskip 0.3\baselineskip}
\def \drule {\vskip 0.4\baselineskip \hrule height.7pt
 \vskip2pt \hrule height.7pt \vskip 0.3\baselineskip}
\tabskip=1em plus 2em minus .5em
\newdimen\digitwidth
\setbox0=\hbox{\rm0}
\digitwidth=\wd0
\catcode`@=\active
\def@{\kern\digitwidth}
%
\def\spose#1{\hbox to 0pt{#1\hss}}
\def\lta{\mathrel{\spose{\lower 3pt\hbox{$\mathchar"218$}}
 \raise 2.0pt\hbox{$\mathchar"13C$}}}
\def\gta{\mathrel{\spose{\lower 3pt\hbox{$\mathchar"218$}}
 \raise 2.0pt\hbox{$\mathchar"13E$}}}
\def\pmb#1{\setbox0=\hbox{#1}%
 \kern-.025em\copy0\kern-\wd0
 \kern.05em\copy0\kern-\wd0
 \kern-.025em\raise.0433em\box0}

\def\sqr#1#2{
	{\vcenter{
		\vbox{
			\dimen1=#1pt
			\dimen2=#2pt
			\dimen11=\dimen1
			\dimen12=\dimen1
			\dimen22=\dimen2
			\divide \dimen22 by -10
			\advance \dimen12 by \dimen22
			\divide \dimen11 by -2
			\dimen3=-0.1\dimen2
			\advance\dimen11 by \dimen3
			\dimen13=\dimen1
			\dimen4=#2pt
			\divide\dimen4 by 10
			\divide\dimen13 by -5
			\kern\dimen13
			\hrule height \dimen4 width #1pt
			\hbox{
				\kern \dimen11
				\vrule width \dimen4 height#1pt
				\kern \dimen12
				\vrule width \dimen4
			}
			\divide\dimen13 by 3
			\kern\dimen13
			\hrule height \dimen4 width #1pt
		}
	}}
}

\def\ueber#1#2{{\setbox0=\hbox{$#1$}%
  \setbox1=\hbox to\wd0{\hss$\scriptscriptstyle #2$\hss}%
  \offinterlineskip
  \vbox{\box1\kern0.4mm\box0}}{}}	

\nopagenumbers
\noindent {\sl Invited talk 
presented at XXXIst Moriond Meeting (held in les Arcs, France in Jan. 1996)
``Dark Matter in Cosmology, Quantum
Measurements, Experimental Gravitation", ed. R. Ansari, Y. Giraud-Heraud \&
J. Tran Thanh Van 
(Gif-sur-Yvette: Eds. Fronti\`eres), in press.}
\vskip -2\baselineskip
\moriondtitle{The DENIS \& 2MASS Near Infrared Surveys}
\moriondtitletwo{and their Applications in Cosmology}
\moriondauthors{Gary MAMON}
\moriondinst{IAP, Paris \& DAEC, Obs. de Paris, Meudon}
\moriondabstract

The DENIS and 2MASS near infrared surveys are presented.
Their 
applications in extragalactic astronomy and cosmology are listed.
The prospects for a rapid spectroscopic followup survey of a near infrared
selected sample of nearly $10^5$ galaxies are illustrated with
Monte-Carlo simulations.

\moriondsec {1. Introduction}

Astronomers have traditionally relied upon optical imagery to view the
Universe, ever since the early introduction of photographic plates, sensitive
to 
blue light.
The main constituents of the Universe in the optical are {\sl
galaxies\/}, and the main constituents of galaxies in the optical are {\sl
stars\/}.
In fact, the visible parts of galaxies contain important amounts of gas and
non-negligible amounts of dust, and perhaps some dark matter.
Moreover, most of the mass in galaxies is thought to be constituted of dark
matter residing in near spherical halos.
Similarly, on the scale of the Universe, most of the baryons are presumed to
be locked up in gas, and most of the mass is believed to be non-baryonic
dark matter.
Working in the optical, one uses galaxies to trace the matter distribution in
the Universe. But galaxies may very well be biased tracers of the underlying
matter content of the Universe.

Near infrared (NIR) light is also sensitive to stars, hence to galaxies.
But there are important differences.

1) NIR light is typically 10 times less extinguished by dust than
optical light. 
This means that the cores of galaxies, hidden in the optical, are visible in
the NIR.
This is illustrated in Figure 1, which shows the same galaxy, in blue ($B$)
and NIR ($K'$) light.
The white pattern cutting through the galaxy in the blue image, is caused by
internal extinction by dust, and is invisible in the NIR image.
The galaxy is much more symmetric in NIR light and the shapes of the
isophotes from disky inside to boxy outside can only be seen on the NIR image.
\moriondpsfigurenr{1}{10.3}{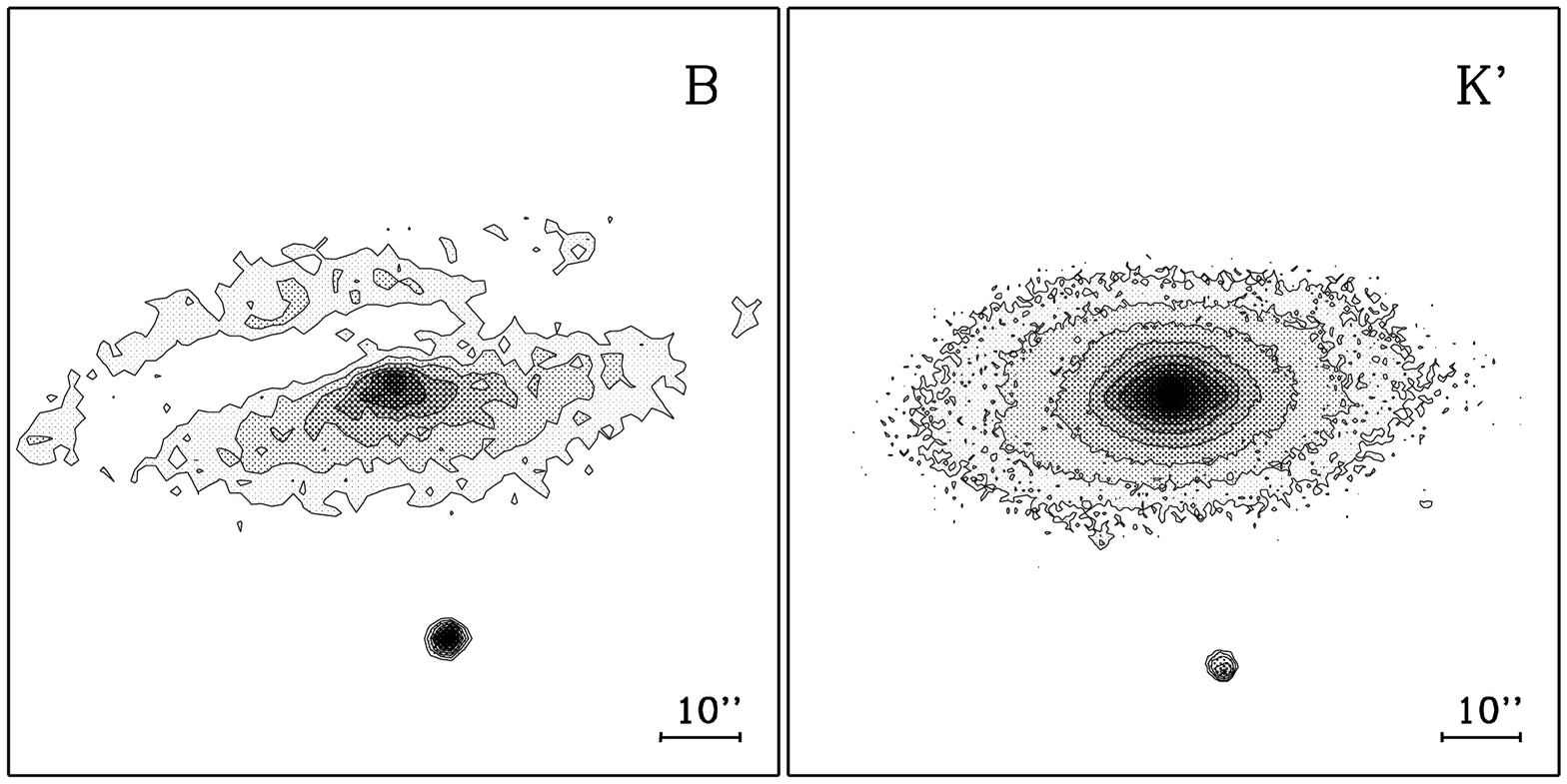}{Galaxy NGC 7172, seen in blue
({\it left\/}) and NIR ({\it right\/}) light.$^{1)}$
Object at bottom is a foreground star.
} 

2) The second difference is that, contrary to NIR light,
optical light is extremely sensitive to
populations of newborn stars, because the more massive (and hot) stars are
extremely 
luminous, and are the only very blue stars (for cooler stars, the blue band is
in the Wien part of the spectrum).
As stars exhaust their nuclear hydrogen and evolve to the red giant branch,
they become substantially redder.
The more massive ones evolve to cool Red Giants in very short times (a few
million 
years) and continue their evolution and finally
blow up as supernovae in even shorter times.
The NIR band picks up principally cool stars, while the hot, massive, and young
ones, moderately visible as the NIR lies in the Rayleigh-Jeans part of
their spectrum, are too rare to contribute significantly to NIR light.
As a result, blue light tends to pick up galaxies with very recent star
formation, while NIR light is less affected by very recent star formation and
traces better the {\it stellar mass\/} content of galaxies.

However, NIR light {\it is\/} affected by star formation occurring roughly 10
million years ago, because after 10 million years the typical very massive
stars are in their Red Giant or subsequent phases before going supernova.
Nevertheless, averaged over all epochs of star formation, the NIR wavelength
domain seems to be the optimal one to avoid the effects of star formation.


These differences between optical light and NIR light have several important
implications:

1) The lack of extinction allows one to probe the large-scale distribution of
galaxies behind the dust-filled plane of our own Galaxy.$^{2)}$

2) The galaxy tracers of the underlying matter distribution in the Universe
is less biased by recent star formation in the NIR than in other bands.

\moriondsec{2. The DENIS and 2MASS Surveys}

With these advantages in mind, two consortia have set out to map the sky at
NIR wavelengths: {\sl DENIS\/} (DEep Near Infrared Southern Sky Survey) and
{\sl 2MASS\/} (Two Micron All Sky Survey)
The characteristics of the two surveys are given in Table 1.
%
%

\medskip

\vbox{
\centerline {\bf Table 1: The DENIS and 2MASS surveys}

\drule
\halign to \hsize{
#\hfil & #\hfil & #\hfil \cr
		& DENIS	& 2MASS \cr
\noalign{\srule}
Institutions	&	Meudon, IAP, Leiden ... 	&	UMass, IPAC,
	CfA ... \cr
Hemispheres	&	South ($\delta < 2^\circ$)	& North + South \cr
Telescopes	&	ESO 1m		&	2 new 1.3m	\cr
Color bands	&	$I\, (0.8\,\mu{\rm m})$, $J\, (1.25\,\mu{\rm m})$, $K_s
\,(2.15\,\mu{\rm m})$ 	&	$J$, $H \,(1.65\,\mu{\rm m})$, $K_s$ \cr 
Detector	&	CCD $1024^2 \,(I)$, NICMOS-3 $256^2\,(JK)$	&
NICMOS-3 $256^2$ \cr
Pixel size	&	$1''$ ($I$), $3''$ ($JK$)	&	$2''$ \cr
Quantum efficiency	&0.4 ($I$), 0.65 ($JK$)	&	0.65 	\cr
Exposure time	&	$8\,{\rm s}\ (I)$, $9\times 1\,{\rm s}\ (JK)$	&
$6 \times 1.3\,\rm s$ \cr
Read-out noise	&	$8\,e^-\ (I)$, $30\,e^-\ (JK)$	&	$30\,e^-$ \cr
Observing mode	&	Stop \& Stare	&	Freeze-Frame Scan \cr
Scan geometry	&	$12' \times 30^\circ$	&	$8' \times 6^\circ$
\cr
Survey years	&	1996--2000	&	1997--2000 (N), 1998--2000
(S) \cr
Primary data	&	4000 GBytes	&	19$\,$000 GBytes \cr
Cost		&	\$3 million	&	\$30 million \cr
}
\srule
\noindent Notes: The NIR exposures are dithered for better angular
resolution.

}

\moriondsec{3. Cosmological Applications}

The applications of NIR surveys such as DENIS and 2MASS
for extragalactic astronomy and cosmology
have been described elsewhere$^{3,4,5,6,7)}$
and
are briefly outlined again here.

\moriondsubsec{3.1 Statistical properties of NIR galaxies\/}

The large sample sizes ($>10^4$, $10^5$, and $10^6$ galaxies in $K$, $J$, and
$I$, respectively, for DENIS$^{5)}$
will help study correlations between properties.

\moriondsubsec{3.2 Two-dimensional structure of the local Universe\/}.

Catalogs of groups and clusters will be obtained from the galaxy lists
extracted from the survey images.
Statistical measures of large-scale structure will be obtained (such as the
angular correlation function and higher order functions, counts in cells, and
topological measures), in particular statistics on the full sphere from 2MASS
data.
Not only are NIR bands cleaner than optical bands to study the large-scale
distribution of galaxies in the Universe (because less sensitive to recent
star formation), but the multi-color aspect of the DENIS and 2MASS surveys
allows one to see how all this 2D structure will vary with
waveband, 
and indicate possible biases when going from one waveband to
another.
The alternative is that any such difference in 2D structure may be a
reflection of selection effects, but we are working hard on avoiding this.
Moreover, NIR surveys will probe the largest local concentration of
matter, the {\sl Great Attractor\/}, which is situated roughly right behind the
Galactic Plane$^{8)}$, and contains the closest rich cluster of
galaxies.$^{9)}$
Unfortunately, although extinction is beaten at low galactic latitudes,
confusion with stars becomes a serious issue, for example on the accuracy of
the photometry of fairly bright galaxies.$^{10)}$

\moriondsubsec{3.3 Color segregation\/}

Instead of studying structure versus waveband, one can study the inverse
problem of understanding colors as a function of structure, hence
environment.
Color segregation is a potentially powerful probe of three dimensional
morphological segregation of galaxies in the Universe ({\it e.g.\/,} the
fact that the
cores of galaxy clusters have the highest fraction of galaxies with
elliptical morphological types).

\moriondsubsec{3.4 Normalization of galaxy counts at the bright-end}

Counting galaxies as a function of apparent magnitude provides better results
at the faint-end than at the bright-end, simply because the bright-end
suffers from very poor statistics (in the local uniform Universe, galaxy
counts rise roughly as ${\rm dex} [0.6\,m]$).
If our Local Group sits in an underdense region, we should see a lack of
galaxies at the very bright end of the galaxy counts, which is brighter than
the DENIS and 2MASS complete/reliable extraction limits.
However, the 
error bars in recent studies$^{11,12,13)}$
are too large to draw firm conclusions, and very wide-angle surveys such
as DENIS and 2MASS will bring them down.

\moriondsubsec{3.5 Mapping interstellar extinction}

Among the many ways one can map the interstellar extinction in our Galaxy,
one is to use galaxy counts$^{14,15)}$, since the count
normalization is shifted downwards when galaxies are
extinguished. 
Galaxy colors may provide better extinction estimates.$^{6)}$

\moriondsubsec{3.6 Cosmic Dipole}

With its full sky coverage, 2MASS will be able to probe the {\sl cosmic
dipole\/}, which computes the vector sum of the flux {\it vectors\/} from the
detected extragalactic objects, and which should be close to the peculiar
acceleration of our galaxy, assuming that galaxy NIR light is a good tracer
of the total mass content of the Universe.
In particular, it will be interesting to compare the cosmic dipole with that
obtained with the sparser IRAS galaxy samples.

\moriondsec{4. A spectroscopic followup}

There is much to be gained from knowing the third dimension in any galaxy
survey. 
In particular, the statistics of the 3D galaxy distribution can be studied,
and projection effects are virtuaally eliminated, 
although the 3D statistics are messed up by peculiar velocities.
3D color segregation can be studied as well as the convergence of the cosmic
dipole with distance.
And finally one gains access to the internal kinematics of structures.

For these reasons, a spectroscopic followup of the DENIS and 2MASS samples
are highly desirable.
In particular, DENIS will extract with high completeness and reliability over
$\simeq 160\,000$ galaxies at $J < 14.4$.$^{5)}$
Only $25\,000$ galaxies are expected in the largest complete and reliable
$K$-band sample$^{4)}$, although if cooling of the DENIS optics is
implemented this 
Autumn, as scheduled, this number could be multiplied by three to four.
The following discussion attempts to optimize the time for obtaining the
largest complete $J$ selected spectroscopic sample.


\moriondsubsec{4.1 Simulated galaxy samples}

To begin, galaxy samples are simulated with random
positions in a uniform Universe, random blue galaxy luminosities from a
Schechter$^{16)}$ 
luminosity function, random bulge/disk ratio, disk inclination, and galactic
latitude, plus
scatter in the surface brightness versus luminosity relations for bulges and
disks. 
Working separately on bulges and disks, the blue and NIR apparent magnitudes
of each component are
estimated using standard colors, 
$k$-corrections (the
effect of redshifting a spectrum through a fixed observation 
wavelength filter), but no luminosity evolution.
Samples of typically $25\,000$ galaxies are simulated with $z < 0.25$ and $L
> 10\,L_*$, and 
subsamples of 
typically a few thousand galaxies are
extracted with apparent magnitude limits in the blue or the NIR.

\moriondsubsec{4.2 Mean galaxy surface brightness within fiber apertures}

The surface brightness of galaxies are well represented by
$\Sigma = \Sigma(0) \exp[(-r/r_1)]^\beta$, where $\beta = 1$ for exponential
disks and $\beta = 1/4$ for $r^{1/4}$ bulges.
The mean surface magnitude within a circular aperture centered on the galaxy
is then 
$$
\langle\mu\rangle = \mu_0 - 2.5 \log \{2[1-(x+1)\exp(-x)]/x^2\} \ ,
\eqno(1)$$
for exponential disks and
$$
\langle\mu\rangle = \mu_0 - 2.5 \log (8 /[b^8  x^2]) - 2.5 \log \gamma(8,
b x^{1/4}) \ ,
\eqno(2)$$
for $r^{1/4}$ bulges.
Here, $b = 7.67$ and
$$
\gamma (8,b x^{1/4}) = 5040 - \exp(-b x^{1/4}) \sum_{k=0}^7
{7!\over k!} (b x^{1/4})^k \ ,
\eqno(3)$$
and, in both cases, $x = \theta_{\rm fib} / (2 \theta_1)$, with the
angular radius 
$\theta_1 = {\rm dex}[-0.2(m-\mu_0)]/c^{1/2}$ is
the angular scale length for exponential disks and
the angular effective radius for $r^{1/4}$ bulges, 
and where $c = 2\pi$ for exponential disks 
and $c = 8!\, \pi / b^8 = 0.0106$ for $r^{1/4}$ bulges.
We correct the mean surface brightness for inclination and internal
extinction of the disk, for Galactic extinction, and for $k$-corrections
and cosmological dimming ($\Sigma \sim (1+z)^{-4}$).

Because spectroscopy of nearby galaxies is done in the optical (we know very
little of galaxy spectra in the NIR), one has to be careful when making
optical observations of a NIR selected sample.
Figure 2 shows the optical surface magnitudes (eqs. [1-3], corrected
as explained above) versus NIR magnitude for 
a $J < 14.1$, subsample and illustrates that the lowest
surface brightness galaxies (which are the most difficult to obtain spectra
for, see below) are distant ellipticals.
\moriondpsfigurenr{2}{12}
{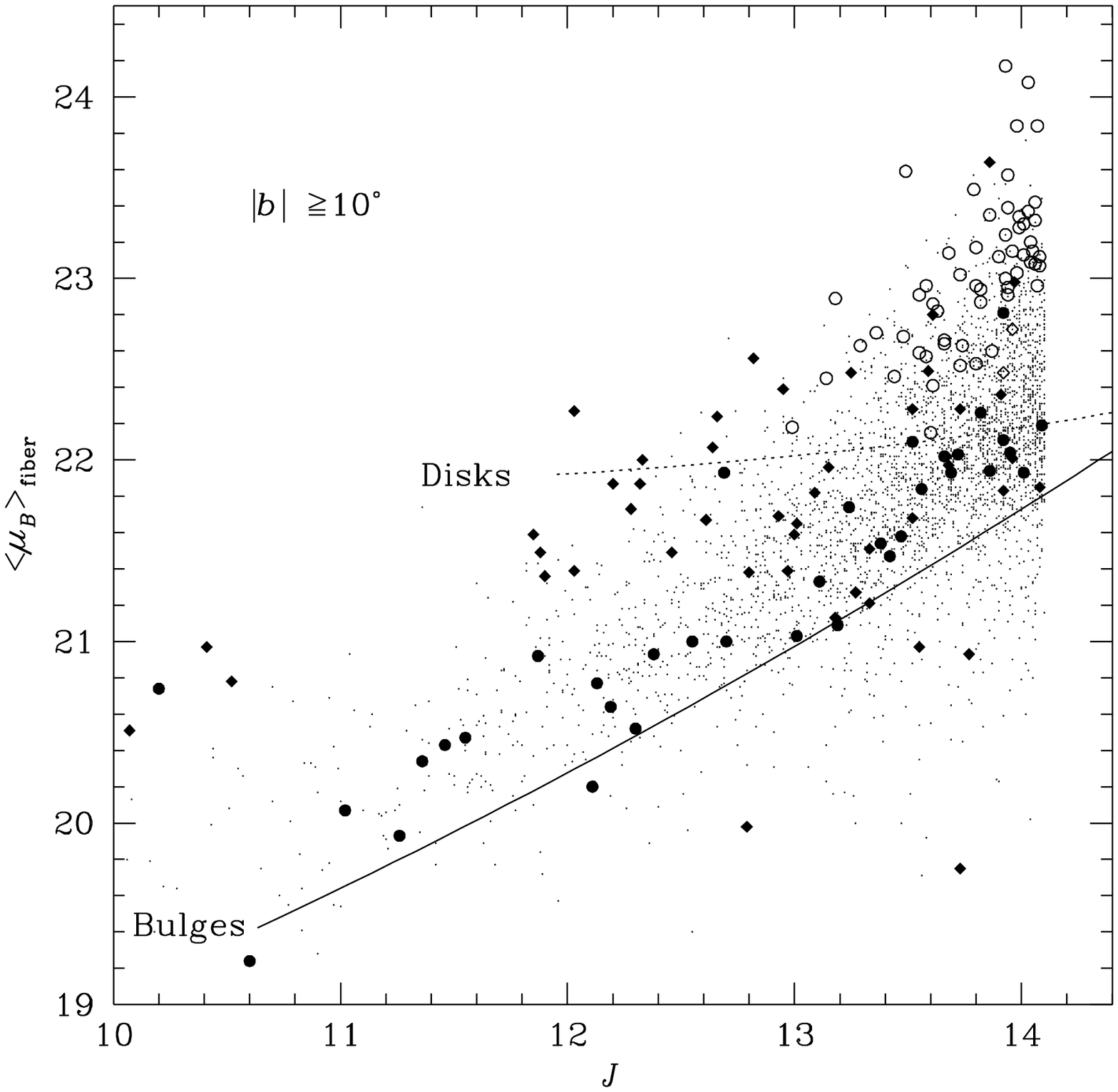}
{Simulation of surface magnitudes of galaxies, averaged over FLAIR $6.7''$
fibers, as 
a function of NIR $J$ magnitude for an NIR selected selected sample of 3300
galaxies, with galactic latitude $|b| > 10^\circ$. 
Late-type spiral galaxies ($\geq 90\%$ of disc)
are shown as {\it diamonds\/} and
elliptical galaxies ($\geq 90\%$ of bulge) by {\it circles\/}.
The {\it full symbols\/} correspond to nearby galaxies ($z \leq
0.03$) and the {\it open symbols\/} to distant galaxies ($z \geq
0.1$).
Galaxies with intermediate bulge/disc ratio, as well as galaxies at
intermediate redshift are shown as {\it points\/}.
The {\it curves\/} represent the theoretical relations for nearby galaxies 
(for bulges we have adopted a luminosity dependent central surface
magnitude$^{17)}$), 
with colors corresponding to a reddening at the galactic pole of 0.17.
Discs and bulges above their respective theoretical curves have greater
reddening, cosmological 
dimming, and 
$k$-correction.
Discs below the theoretical disc curve are inclined.
}

\moriondsubsec{4.3 Time constraints for different instruments}

The number of spectroscopic nights required to
achieve a complete $J$-limited sample in the southern hemisphere has been
estimated for different telescopes in the Southern hemisphere with additional
Monte-Carlo simulations.
The mean surface
magnitudes of galaxies, 
computed over different fiber sizes, are converted to observing
times for continuum $S/N = 5$, {\it assuming that all telescopes have the
same transmission from sky to detector\/}, and normalizing to the
published$^{18)}$ performance of FLAIR-II, a 92-fiber spectroscope on the
very wide-field ($6^\circ$ Schmidt) UKST telescope in Australia.
The maximum time to reach $S/N = 5$ is recorded for each set of $N_f$
galaxies (where 
$N_f$ is the number of fibers), after the galaxies have been first sorted by
galactic latitude.

Figure 3 shows the comparison of the different available (or soon to be) or
potential (FIFI, FLAIR IIIa,b,c) instruments.
Only 60 to 85 nights are required for an $80\,000$ galaxy sample limited to
$J < 13.7$ on an upgrade of FLAIR II, with more fibers and automatic
configuration.
This would correspond to one year at 50\% pressure on the UKST telescope
(which is also used for Schmidt plate photographic surveys).
\moriondpsfigurenr{3}{12}
{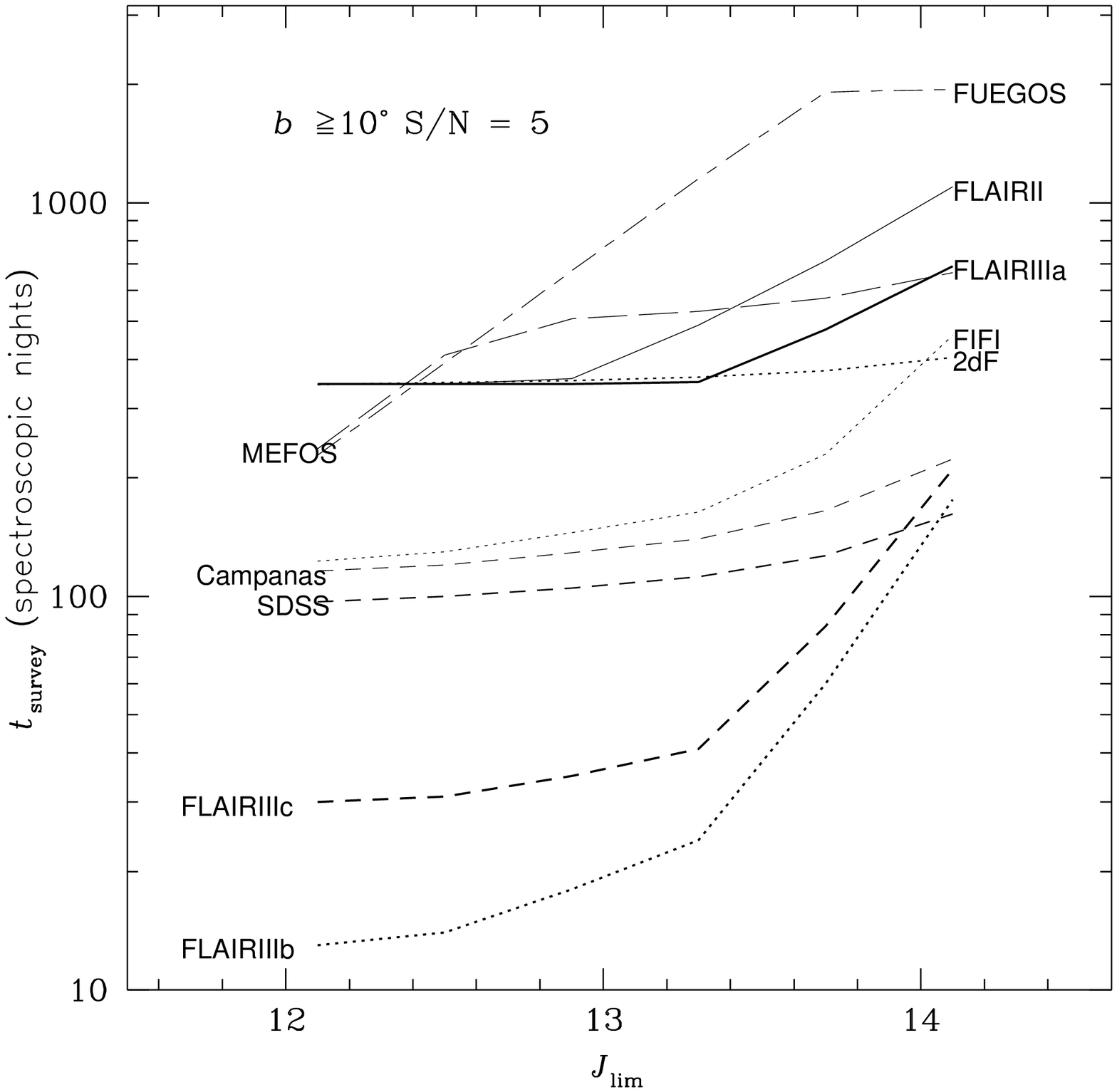}
{Estimates of time in spectroscopic nights to complete 
{\it without loss of galaxies\/} 
a survey with $|b| \geq 10^\circ$ and $S/N \geq 5$, using mean surface
brightnesses within fibers, for the simulated galaxy sample.
FLAIR IIIa would have
150 fibers, with slow day-time configuration.
FLAIR IIIb and IIIc would have 250 fibers with respectively 30 and 10 minute
on-telescope configuration. 
FIFI is a copy of MEFOS that could easily be 
mounted at the SAAO in South Africa.
}

Discussions are currently underway for upgrading FLAIR II and performing a
spectroscopic followup to DENIS. These involve Q. Parker and M. Colless in
Australia, W. Saunders in Edinburgh, and V. Cayatte, H. Di Nella,
R. Kraan-Korteweg, G. Paturel and others in France.
Our target is to complete the spectroscopic survey in 2000 or 2001, starting
towards the end of the DENIS survey.


\moriondrefhead

\moriondpaper{H\'eraudeau, P., Simien, F. \& Mamon,
G.A.}{1996}{A\&AS}{117}{417}{1}

\moriondproceed{Mamon, G.A.}{1994}{``Unveiling  Large-Scale Structures behind
the Galactic Plane'', ASP vol. 67}{C. Balkowski \& R.C. Kraan-Korteweg}{San
Francisco}{ASP}{53}{2}

\moriondpaper{Mamon, G.A.}{1994}{ApSS}{217}{237}{3}

\moriondproceed{Mamon, G.A.}{1995}{35th Herstmonceaux Conference:
``Wide-Field Spectroscopy and the Distant Universe''}{S.J. Maddox \&
A. Arag\'on-Salamanca}{Singapore}{World Scientific}{73}{4}

\moriondproceed{Mamon, G.A.}{1996}{``Spiral Galaxies in the Near-IR''}
{D. Minniti \& H.-W. Rix}{Berlin}{Springer}{195}{5}

\moriondproceedinpress{Mamon, G.A., Banchet, V., Tricottet, M. \& Katz, D.}
{1997}{``The Impact of Large Scale Near
Infrared Surveys''}{F. Garz\'on}{Dordrecht}{Kluwer}{6}

\moriondproceedinpress{Schneider, S.}{1997}{``The Impact of Large Scale Near
Infrared Surveys''}{F. Garz\'on}{Dordrecht}{Kluwer}{7}

\moriondpaper{Kolatt, T., Dekel, A. \& Lahav}{1995}{MNRAS}{275}{797}{8}

\moriondpaper{Kraan-Korteweg, R.C., Woudt, P.A., Cayatte, V., Fairall, A.P.,
Balkowski, C. \& Henning, P.A.}{1996}{Nature}{379}{519}{9}

\moriondproceedinpress{Chester, T.}{1997}{``The Impact of Large Scale Near
Infrared Surveys''}{F. Garz\'on}{Dordrecht}{Kluwer}{10}

\moriondpaper{Mobasher, B., Sharples, R.M. \& Ellis,
R.S.}{1986}{MNRAS}{223}{11}{11}

\moriondinpress{Gardner, J.P., Sharples, R.M., Carrasco, B.E. \& Frenk,
C.S.}{1996}{MNRAS}{12}

\moriondsubmit{Huang, J.-S., Cowie, L.L., Ellis, R.S., Gardner, J.P.,
Glazebrook, K., Hu, E.M., Songaila, A. \& Wainscoat, R.J.}{1996}{ApJ}{13}

\moriondpaper{Burstein, D. \& Heiles, C.}{1978}{Ap. Lett.}{19}{69}{14}

\moriondpaper{Burstein, D. \& Heiles, C.}{1982}{AJ}{87}{1165}{15}

\moriondpaper{Schechter, P.L.}{1976}{ApJ}{203}{297}{16}

\moriondpaper{Sandage, A. \& Perelmuter, J.-M.}{1990}{ApJ}{361}{1}{17}

\moriondpaper{Parker, Q.A.}{1995}{Spectrum}{7}{17}{18}

\bye